\documentclass[aps,pra,10pt,twocolumn,titlepage,showkeys,nofootinbib,showpacs,superscriptaddress]{revtex4-1}

\usepackage{graphicx}
\usepackage{url}
\usepackage{amsmath}
\usepackage{amssymb}
\usepackage{bm}
\usepackage{dsfont}
\usepackage{calc}
\usepackage{array}
\usepackage{courier}
\usepackage{scalefnt}
\usepackage{braket}
\usepackage{times}
\usepackage{units}
\usepackage{yfonts}
\usepackage{calc}
\usepackage{color}
\usepackage[normalem]{ulem} 

\newcommand{\mF}{m_\mathrm{F}}

\newcommand{\A}{\mathrm{A}}
\newcommand{\B}{\mathrm{B}}
\newcommand{\e}[1]{\mathrm{e}^{#1}}
\newcommand{\sg}{{\mathrm{s}_\mathrm{g}}}
\newcommand{\su}{{\mathrm{s}_\mathrm{u}}}
\newcommand{\ER}{E_\mathrm{R}}
\newcommand{\n}{{(n)}}
\newcommand{\Ns}{N_\text{s}}
\newcommand{\JBC}{{\Delta J}}

\newcommand{\as}{a_\mathrm{s}}
\newcommand{\expect}[1]{\langle #1 \rangle}
\newcommand{\bd}[1]{\hat{b}_{#1}^{\dagger}}
\newcommand{\dx}{\int\! d^3{x}\ }
\newcommand{\BHM}{\mathrm{BH}}
\newcommand{\EBHM}{\mathrm{EBH}}
\newcommand{\dit}{density-induced tunneling }
\newcommand{\di}{density-induced }
\newcommand{\dd}{next-neighbor }
\newcommand{\fig}[1]{Fig.~\ref{#1}}
\renewcommand{\sec}[1]{Sec.~\ref{#1}}
\newcommand*\phantomas[3][c]{\ifmmode \makebox[\widthof{$#2$}][#1]{$#3$}\else \makebox[\widthof{#2}][#1]{#3}\fi}

\begin{document}

\title{Quantum phases in tunable state-dependent hexagonal optical lattices}

\author{Dirk-S\"oren L\"uhmann}
\author{Ole J\"urgensen}
\author{Malte Weinberg}
\author{Juliette Simonet}
\author{Parvis Soltan-Panahi}
\author{Klaus Sengstock}

\affiliation{Institut f\"ur Laser-Physik, Universit\"at Hamburg, Luruper Chaussee 149, 22761 Hamburg, Germany}


\begin{abstract}
We study the ground-state properties of ultracold bosonic atoms in a state-dependent graphene-like honeycomb optical lattice, where the degeneracy between the two triangular sublattices A and B can be lifted. We discuss the various geometries accessible with this lattice setup and present a novel scheme to control the energy offset with external magnetic fields. The competition of the on-site interaction with the offset energy leads to Mott phases characterized by population imbalances between the sublattices. For the definition of an optimal Hubbard model, we demonstrate a scheme that allows for the efficient computation of Wannier functions. Using a cluster mean-field method, we compute the phase diagrams and provide a universal representation for arbitrary energy offsets. We find good agreement with the experimental data for the superfluid to Mott insulator transition.
\end{abstract}


\maketitle

Ultracold quantum gases in optical lattices perform impressively well in simulating condensed matter Hamiltonians and allow for the realization of fully novel quantum systems. While seminal experiments have been performed in square and cubic lattices, the recent development of non-cubic optical lattice geometries leads to a variety of new possibilities. Recent achievements include the triangular \cite{Becker2010,Struck2011} and  honeycomb  \cite{SoltanPanahi2011,SoltanPanahi2012} optical lattices as well as checkerboard \cite{Sebby-Strabley2006, Wirth2011}, quasi-honeycomb \cite{Tarruell2012,Uehlinger2013} and Kagom\'e \cite{Jo2012} systems. Quantum phases in superlattices have been theoretically studied in one-dimensional lattices \cite{Buonsante2004,Buonsante2004b,Rousseau2006}, special two-dimensional cases \cite{Buonsante2005} and, employing mean-field methods, in higher dimensions \cite{Chen2010}, to name only a few examples.

In solid-state physics, graphene is a prominent example for the honeycomb lattice \cite{Geim2007}. Here, the linear dispersion relation at the Dirac points  gives rise to phenomena such as quasi-relativistic particles and an anomalous quantum Hall effect \cite{CastroNeto2009}. This topological peculiarity has drawn much attention to experiments with ultracold atoms in  honeycomb  optical lattices (see Refs.~\cite{Polini2013,Zhang2012,Zhu2007,Wu2008,Chen2011,Lim2012}). For bosonic atoms, the superfluid to Mott-insulator transition could be observed for single- and multi-component quantum gases in optical lattices \cite{SoltanPanahi2011}, while fermionic atoms have been used to study the dispersion relation \cite{Tarruell2012}. In spin mixtures, a twisted-phase superfluid has been observed resulting from an unconventional hybridization of the lowest bands has been observed in quantum mixtures \cite{SoltanPanahi2012}.

In this paper we discuss various novel schemes of tuning hexagonal lattice structures allowing for the realization of new quantum phases for different spin states. The lattice potential arises here from the interplay between a scalar light shift and a non-negligible vectorial light shift, involving the light polarization and internal state of the atom. Depending on the polarization of the lattice laser beams, it is possible to create a number of potentials, such as a scalar triangular lattice as well as a pure polarization lattice without intensity modulation. A particularly interesting configuration is that of a honeycomb potential with an additional state-dependent superlattice structure \cite{SoltanPanahi2011}.

\begin{figure*}[t]
\includegraphics{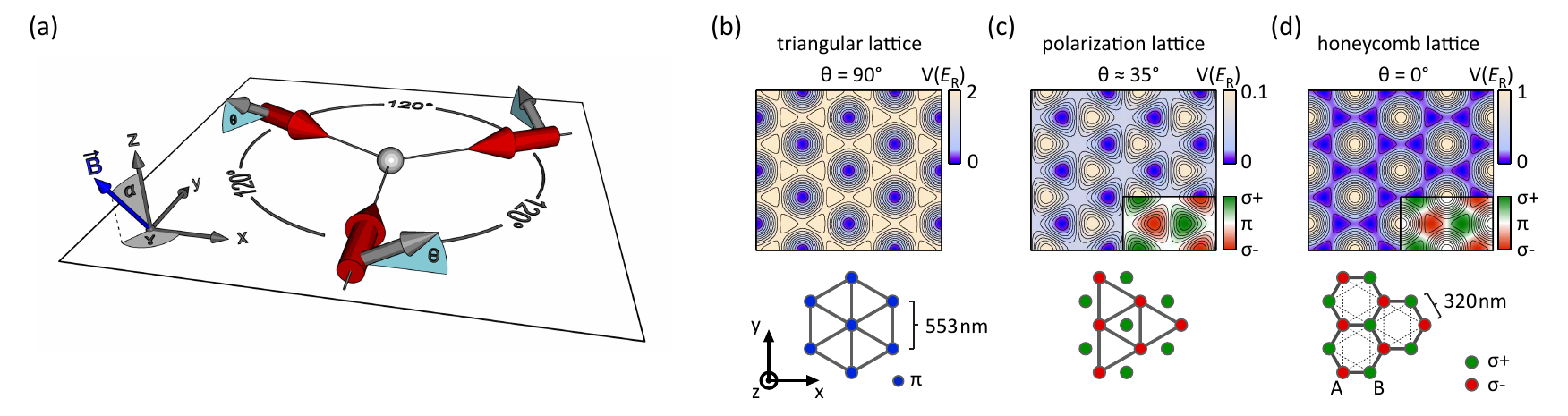}
\caption{(Color online) Setup and possible geometries of versatile spin-dependent hexagonal lattices. (a) Setup of the three running laser beams in the $xy$-plane. The quantization axis of the system is defined by a homogeneous magnetic field \textbf{B}. Its orientation can be quantified by the Euler angles $\alpha$ and $\gamma$. The linear polarization of the three beams encloses an angle $\theta$ with the lattice plane. (b) For $\theta=\pi/2$ (polarization perpendicular to lattice plane) the resulting potential is a triangular lattice with $\pi$-polarized light. The light intensity is strongly modulated. (c) Purely state-dependent polarization lattice in the case of $\theta = \arccos(1/3)/2$, depicted here for atoms trapped at $\sigma^-$ sites and $\alpha=0$. Note that compared with (b) a much weaker confinement arises. (d) With $\theta=0$ the potential forms a spin-dependent honeycomb lattice. The alternating circular polarization lifts the degeneracy of the sublattice sites A and B ($\alpha=0$). The lattice spacing (533\,nm and 320\,nm) are exemplary given for a laser wavelength of $\lambda_L=830\,\text{nm}$.}
\label{Figure01}
\end{figure*}

For the case of the spin-dependent  honeycomb lattice potential, the ground state properties of bosonic gases are investigated in detail. We propose a novel scheme to dynamically control the geometry of the lattice potential that does not require additional light fields in the experimental setup. We demonstrate that this system allows for tailoring Hubbard models with a tunable energy offset between the two different sublattices A and B. This offset is introduced via rotation of the quantization axis of the system defined by an external magnetic field, thus modifying the amplitude of the polarization-dependent light shift. The resulting two-atomic unit cell with non-equivalent sites leads to a series of superfluid and Mott-insulating quantum phases with integer and half integer fillings. Applying a bosonic cluster mean-field method \cite{Buonsante2004b, Jain2004, Buonsante2005b, Hen2009, Hen2010, Pisarski2011, McIntosh2012, Yamamoto2012, Yamamoto2012b, Luhmann2013}, we derive accurate phase diagrams for arbitrary offset energies. The deviations due to off-site processes and next-nearest neighbor tunneling are discussed for the experimental parameters. Furthermore, we derive the Wannier functions for the composite bands of the honeycomb lattice using a localization criterion for an \textit{optimal} Hubbard model. This criterion minimizes the neglected beyond-Hubbard processes and thereby leads to optimal Hubbard parameters. An effortless construction scheme for Wannier functions depending on only one variational parameter is demonstrated. Finally, we show that the accurate calculation of both Wannier functions and phase diagrams leads to results that agree well with the experimental results in Ref.\,\cite{SoltanPanahi2011}.

We start with a description of a very flexible spin-dependent graphene-like optical lattice setup. In the second section, we focus on the case of the honeycomb lattice and discuss the band structure and the construction of Wannier functions. As a consequence of the polarization-dependent light potential, all parameters depend on the magnetic quantum number $\mF$. The Hubbard parameters for tunneling and on-site interaction are derived from the Wannier functions. In addition, also processes beyond the standard Hubbard model, such as next-nearest-neighbor and \di tunneling \cite{Luhmann2012,Jurgensen2012}, are taken into account. In the third section, the phase diagram of bosonic atoms is discussed focusing on the superfluid to Mott insulator transition.  For the honeycomb lattice, conventional mean-field methods suffer from the small number of nearest neighbors \cite{Teichmann2010, Luhmann2013}. Here, the cluster Gutzwiller method is applied that allows to calculate the transition accurately \cite{Buonsante2004b, Buonsante2005b, McIntosh2012, Pisarski2011, Jain2004, Yamamoto2012, Yamamoto2012b, Hen2009, Hen2010, Luhmann2013}. We show that the critical point for integer and half-integer Mott phases depends crucially on an \textit{effective} magnetic quantum number, ranging continuously from $-\mF$ to $+\mF$.

\section{Tunable hexagonal lattices}

Optical lattices are commonly created by the interference of counter-propagating laser beams that form standing waves with defined knots at retro-reflecting mirrors. In contrast, the tunable optical lattices described here are generated by superimposing three traveling waves which intersect in the $xy$-plane at angles of $120^\circ$ as depicted in Fig.\,\ref{Figure01}a \cite{Petsas1994}.  The corresponding wave vectors of the beams are $\mathbf{k}_1=2\pi(0,1,0)/\lambda_L$, $\mathbf{k}_2=\pi(\smash{\sqrt{3}},-1,0)/\lambda_L$ and $\mathbf{k}_3=\pi(-\smash{\sqrt{3}},-1,0)/\lambda_L$. The recoil energy is given by $\ER=h^2/(2m\lambda_L^2)$, where $m$ is the atomic mass. Throughout the paper we use as an example ultracold $^{87}\text{Rb}$ atoms and a laser wavelength of $\lambda_L=830\,\text{nm}$ as in Ref.~\cite{SoltanPanahi2011}. For this wavelength the lattice laser detuning relative to the atomic transitions is still of the order of the fine structure splitting. In addition to the intensity modulation $V_\text{int}(\mathbf{x})$ of the resulting light field, this gives rise to a reasonably strong polarization-induced Stark shift of the magnetic Zeeman substates $|\mathrm{F},\mF\rangle$, which adds to the intensity modulation. The total optical potential can be expressed as a sum of a state-independent and a state-dependent part:
\begin{equation}
	V(\mathbf{x})= -V_0\big[V_{\text{int}}(\mathbf{x}) + V_{\text{pol}}(\mathbf{x})\big].
\end{equation}
$V_0$ denotes the corresponding lattice depth created by two equivalent counter-propagating laser beams \cite{Becker2010,SoltanPanahi2011}. Considering all beams being linearly polarized at an angle $\theta$ with respect to the $xy$-plane (see Fig.\,\ref{Figure01}a), the state-independent potential reads
\begin{equation}\label{eq:IntPotential}
V_\text{int}(\mathbf{x}) = 6+\big[ 1-3\cos(2\theta) \big]\sum_i\cos(\mathbf{b}_i\mathbf{x}),
\end{equation}
where each two reciprocal lattice vectors $\mathbf{b}_i = \varepsilon_{ijk}(\mathbf{k}_j - \mathbf{k}_k)$, span the reciprocal Bravais lattice. The state-dependent part of the optical potential can be obtained by calculating the projection of the light field onto the polarization basis vectors $\boldsymbol\varepsilon_{\mathcal{P}}$, with polarization $\mathcal{P}=\{\pi,\sigma^+,\sigma^-\}$. This basis is determined by the orientation of the system's quantization axis, which can be easily controlled in experiment by a homogeneous magnetic field. In case of the quantization axis pointing along the $z$-axis, the $\boldsymbol\varepsilon_{\mathcal{P}}$ are the (three-dimensional) Jones vectors $\boldsymbol\varepsilon_\pi=(0,0,1)$ and $\boldsymbol\varepsilon_{\sigma^\pm}=(1,\pm \mathrm{i},0)/\smash{\sqrt{2}}$. For an arbitrary orientation of the quantization axis, the basis has to be transformed, such that $\boldsymbol\varepsilon_\pi$ remains parallel to the quantization axis: $\boldsymbol\varepsilon_{\mathcal{P}} \rightarrow R_z(\gamma) R_x(\beta) R_y(\alpha) \boldsymbol\varepsilon_{\mathcal{P}}$. Here, $\alpha, \beta, \gamma$ denote the Euler angles, defining the orientation of the quantization axis and the $R_i$ are the Cartesian rotation matrices. Without loss of generality, we restrict the following considerations to $\beta=0$. The general form of the resulting state-dependent part of the potential reads
\begin{equation}
\label{eq:PolPotential}
V_\text{pol}(\mathbf{x}) = \sqrt{3}(-1)^{\mathrm{F}}\mF\eta\cos(\theta)\sum_iC_i\sin(\mathbf{b}_i\mathbf{x}),
\end{equation}
where the coefficients $C_i$ are given by
\begin{alignat}{5}
\label{eq:C1}
&C_1&&=\cos\theta\cos\alpha - 2&&\sin\theta\sin\alpha\cos\gamma\\
&C_{2,3}&&=\cos\theta\cos\alpha \phantomas[r]{-2}{+}&&\sin\theta\sin\alpha\big[\cos\gamma\pm\sqrt{3}\sin\gamma\big].\label{eq:C2}
\end{alignat}
The dimensionless proportionality factor $\eta = 0.13$ is solely determined by the detuning of the lattice laser (for, e.g. $\lambda_L=1064\,\mathrm{nm}$, $\eta = 0.03$).

Equations \eqref{eq:IntPotential} and \eqref{eq:PolPotential} point out the central role of the lattice beam polarization angle $\theta$. Three fundamentally different scenarios will be discussed in the following. For $\theta = 90^\circ$ the state-dependent part of the potential vanishes. The remaining state-independent potential forms a triangular lattice with deep confinement at each lattice site due to a strong intensity modulation \cite{Becker2010}. As depicted in Fig.\,\ref{Figure01}b, the polarization throughout this system is $\pi$ and the orientation of the quantization axis has no effect on the potential.

\begin{figure}
\includegraphics{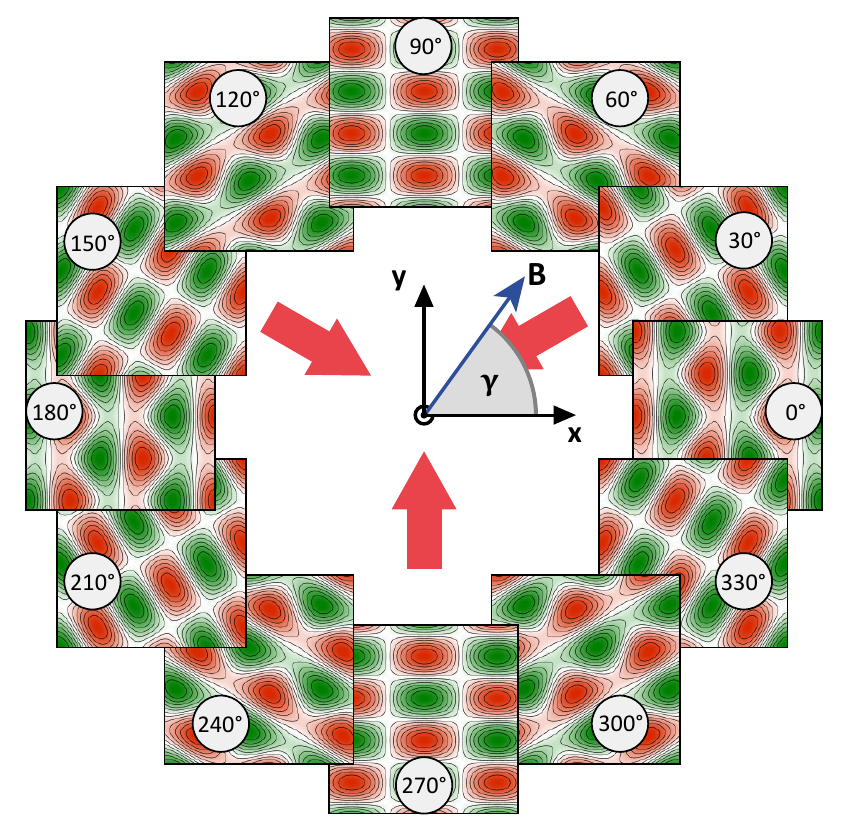}
\caption{(Color online) Stirring of lattice wells in the polarization lattice by rotating the quantization field. In the case of a purely state-dependent potential it is possible to stir $\sigma^+$ and $\sigma^-$ lattice sites around each other by rotating the quantization axis in the lattice plane (i.e. for $\alpha = \pi/2$).}
\label{Figure02}
\end{figure}

In contrast, for $\theta=\arccos(1/3)/2\approx 35.3^\circ$, the modulation of the state-independent potential $V_\mathrm{int}$ vanishes completely, leaving only a constant energy offset and a weak, state-dependent potential (see Fig.\,\ref{Figure01}c). In other words, atoms with $\mF\neq0$ are either trapped on $\sigma^+$ or $\sigma^-$ sites, while atoms with $\mF=0$ remain untrapped. In a spin-mixture of atoms with $\mF \neq 0$ and $\mF = 0$, the periodically modulated density distribution of the atoms with $\mF \neq 0$ presents an interaction lattice for those with $\mF=0$ \cite{McKay2010}. The orientation of the quantization field has a strong impact on the geometry of such a system: rotating the quantization axis from $\alpha=0^{\circ}$ towards the lattice plane at $\alpha = 90^{\circ}$ shifts the lattice basis vectors. 

Such a configuration allows for the realization of a microscopic stack of \emph{stirring spoons}, each covering the area of only one lattice plaquette. Indeed, as depicted in Fig.\,\ref{Figure02}, a rotation of the quantization axis in the lattice plane by the angle $\gamma$ leads to a rotation of the lattice wells around each other. By emulating the effect of the Lorentz force, this stirring offers a novel scheme to create artificial gauge fields. This rotation could also be applied to a spin mixture in order to study interaction induced momentum exchange. Atoms with $\mF\neq0$, trapped in the rotating polarization lattice, shall induce a rotating interaction lattice onto the $\mF=0$ atoms, leading to a transfer of vortices.

In analogy to the purely state-dependent potential, the honeycomb lattice depicted in Fig.\,\ref{Figure01}d, created for $\theta=0^{\circ}$ and $\alpha = 0^{\circ}$, exhibits an alternating pattern of circular polarization. While the potential becomes very large in the center of each hexagon, tunneling processes take place in the shallow channel structure connecting nearest-neighbor lattice sites with contributions of next-nearest neighbor tunneling within each sublattice (see the thin dashed lines in Fig.\,\ref{Figure01}d). Despite the still relatively small proportionality factor $\eta$, this occurrence of only small potential barriers between nearest-neighbor lattice sites causes the state-dependent part of the potential to lift the degeneracy of the two fold atomic basis for atoms with non-vanishing magnetic quantum number. Atoms seeking $\sigma^+$ light are predominantly trapped at the sublattice A, while $\sigma^-$ seeking states occupy the sublattice B and atoms with zero magnetic quantum number still experience a fully symmetric honeycomb potential \cite{SoltanPanahi2011,SoltanPanahi2012}.

\begin{figure*}[t]
\includegraphics{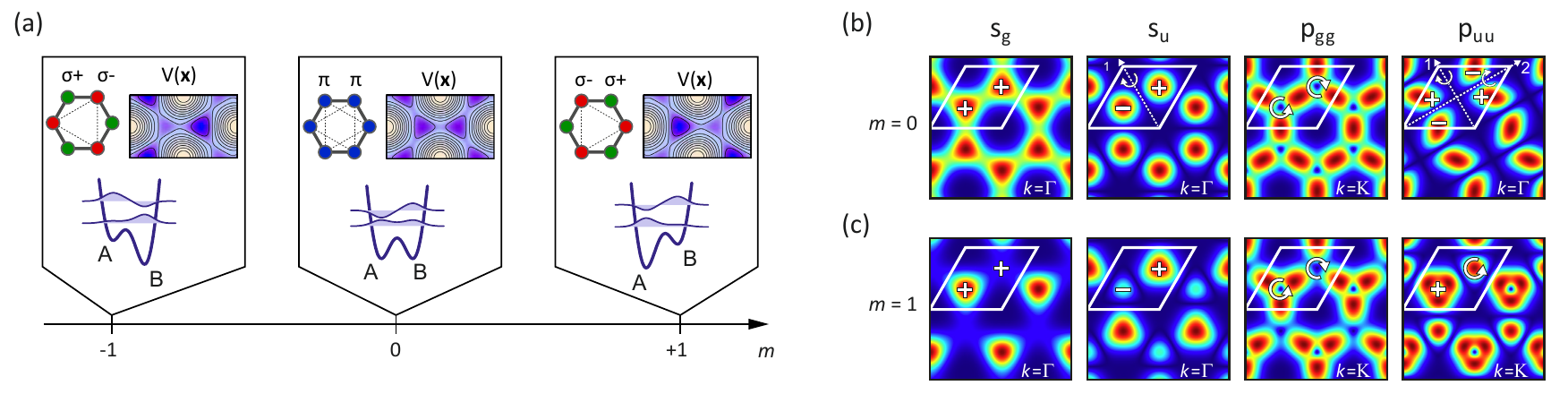}
\caption{(Color online) Band structure of the tunable honeycomb lattice.
(\textbf{a}) The rotation of magnetic field axis with respect to the lattice plane alters the projection of the light field onto the atomic spin-states. Thus, the energy offset $\epsilon$ between the sublattice sites A and B can be continuously adjusted by the Euler angle $\alpha$ in the honeycomb lattice system. This behavior can be described by an effective magnetic quantum number $m$. (\textbf{b},\textbf{c}) Density of the Bloch functions $\ket{\phi^\n_{\mathbf{k}}}$ for the four lowest bands plotted for the momentum $\mathbf{k}=\Gamma$ or K (see Fig.\,\ref{fig:bandstructure}). The results are shown for a lattice depth of $V_0=3\ER$ and an effective magnetic quantum number (\textbf{b}) $m=0$  and (\textbf{c}) $m=1$, where the deep blue color corresponds to nodes in the wave function with zero density. Indicated in white are symmetry axes and the sign of the wave function, where the circular arrows indicate a complex value with a rotating complex phase.}
\label{Waves}
\end{figure*}

As a central aspect here, the control over the quantization axis allows for a \emph{continuous} adjustment of the symmetry of the potential from triangular to honeycomb. By rotating the quantization field into the lattice plane, the projection of the circular polarization onto the atomic spin-states vanishes in compliance with equations (\ref{eq:C1}) and (\ref{eq:C2}), which truncate to $C_{1,2,3}=\cos\alpha$. Rotating the magnetic field beyond the lattice plane, or $\alpha>90^{\circ}$, leads to an exchanged pattern of circular polarizations and, thus, a deeper trapping on the respectively other sublattice as depicted in Fig.\,\ref{Waves}a.

This behavior is equivalent to the realization of an \emph{effective} magnetic quantum number
\begin{equation}
	m=(-1)^{\mathrm{F}+1}\mF\cos\alpha
	\label{eq:m}
\end{equation}
of which we make use in the following calculations. Here, the sign-term used earlier in equation (\ref{eq:PolPotential}) accounts for the respective Land\'e g-factor. Thus, scaling the effective magnetic quantum number $m$ is sufficient to transfer the presented results to an arbitrary atomic species and detuning. All results presented in the following will be focused on this case of the state-dependent honeycomb lattice ($\theta=0$).
 
In this part, we have discussed a versatile state-dependent optical lattice setup. The orientations of the polarization vector and of the quantization field allow to realize various lattice geometries, which can be dynamically modified. This gives access to novel experimental schemes such as the creation of stirring patterns or interaction lattices. In the following the special case of the honeycomb lattice with tunable offset energy is investigated in detail.

\section{Band structure and Wannier functions} \label{sec:BandStructureAndWannierFunctions}
\subsection{Band structure and Bloch functions}

The band structure and the Bloch functions are obtained by diagonalizing the single-particle Hamiltonian $\hat H_0=\mathbf{p}^2/2m+V(\mathbf{x})$, where $m$ is the mass of the atoms and $V(\mathbf{x})$ is the non-separable two-dimensional honeycomb potential for $\theta=0$. The solutions are found by expanding both the periodic lattice potential and the wave functions in two-dimensional plane waves $c_{k_a,k_b} \exp(\mathrm{i}k_a a + \mathrm{i}k_b b)$, spanned by the two lattice vectors $a$ and $b$ and the respective quasi-momenta $k_a$ and $k_b$. Applying the Bloch theorem, the Schr\"odinger equation can be solved for all quasi-momenta  $\mathbf{k}=(k_a,k_b)$. The four lowest bands of the band structure are depicted in \fig{Bands} for various values of $m$. For $m=0$, the two lowest bands $\sg$ and $\su$ show the typical back-folded band structure of a honeycomb lattice. The respective Bloch waves $\ket{\phi^{(n)}_\mathbf{k}}$ (\fig{Waves}b) are even and odd combinations of s-wave-type solutions of the individual sublattices A and B. The relative weight of both solutions depends on the effective magnetic quantum number $m$ and is equal for $m=0$. In the latter case, the inversion symmetry is reflected by the existence of Dirac cones at $\mathbf{k}=\text{K}$ corresponding to massless Dirac particles as in graphene \cite{Zhu2007,Chen2011,Zhang2012,Tarruell2012}. When increasing $m$ the symmetry is lifted introducing an energy offset $\epsilon$ between the minima of the sublattices. As a consequence, a gap opens between the Dirac cones (\fig{Bands}b) and the Dirac particles obtain a finite mass. Therefore, the lattice offers the possibility to study the continuous transition from a graphene-like lattice to a gapped band structure, which is well accessible experimentally. For $m=1$ the site offset is rather large and the densities of $\ket{\phi^{(\sg)}_\mathbf{k}}$ and $\ket{\phi^{(\su)}_\mathbf{k}}$ differ strongly on A and B sites (\fig{Waves}c).

Figure \ref{Bands} reveals another Dirac point connecting the fourth and fifth band \cite{Wu2008}. A special feature of the third band is the extremely flat energy dispersion allowing for Wigner crystallization \cite{Wu2007,Wu2008}.
In the following, we will focus on the lower two bands and neglect the occupation of higher bands.

\subsection{Definition of Wannier functions}
\begin{figure}[t]
\includegraphics{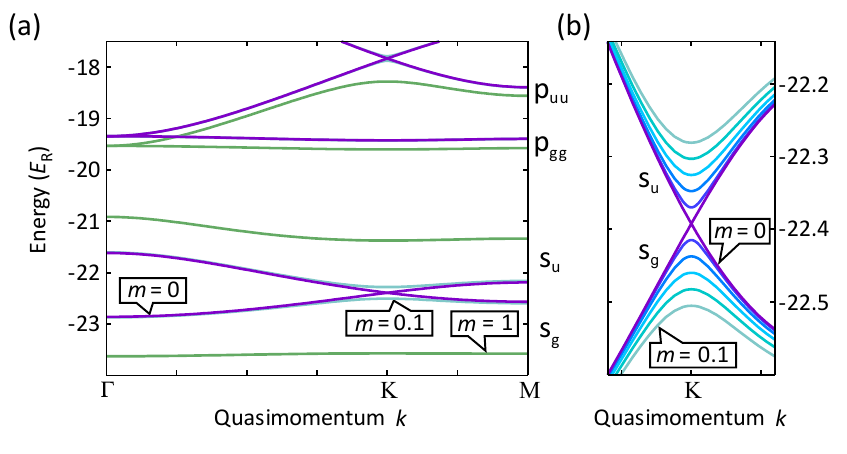}
\caption{(Color online) (\textbf{a}) Band structure  $E^\n_{\mathbf{k}}$ showing the four lowest bands at a lattice depth $V_0=3\ER$  for
$m=0$, $0.1$, $1$ in unit of the recoil energy $\ER$. (\textbf{b}) Close-up of Dirac cones  for $m=0$, $0.02$, ..., $0.1$.}
\label{sec:DefWannier}\label{fig:bandstructure}\label{Bands}
\end{figure}

Since the unit cell of the honeycomb lattice has two lattices sites, the definition of Wannier functions $\ket{w_\A}$ and $\ket{w_\B}$ for the respective sublattices is in general not straight forward. However, we show in the following that for the spin-independent case $m=0$ with equal sublattices, the Wannier problem is solved directly by an equal superposition of $\ket{\phi^{(\sg)}_\mathbf{k}}$ and $\ket{\phi^{(\su)}_\mathbf{k}}$ bands.
Recently, the Wannier functions have been calculated for honeycomb optical lattices using the Marzari-Vanderbilt method in Ref.~\cite{Walters2013} and in Ref.~\cite{IbanezAzpiroz2013,IbanezAzpiroz2013b} for the tight-binding approximation as well as in Ref.~\cite{Uehlinger2013} using the eigenstates of band-projected position operators.

The general goal of the definition of Wannier functions is to provide a Hubbard model that is well suited for the description of the many-body problem. For this purpose, the amplitudes of the leading-order neglected processes must be negligibly small. This includes, e.g., off-site interactions, next-nearest neighbor tunneling and density-induced tunneling. Maximally localized generalized Wannier functions for lattices with multi-atomic unit cells are highly non-unique. In fact, for the honeycomb lattice, an infinite number of orthonormal basis sets exist that are more or less well localized on individual lattice sites. The usual attempt is to minimize the so-called \textit{spread function} \cite{Marzari1997, Marzari2012, Walters2013,IbanezAzpiroz2013,IbanezAzpiroz2013b} in order to define Wannier functions localized to individual lattice sites. However, it is a priori not clear, that these Wannier functions also lead to a Hubbard model that is the best possible description of the system. This is of particular importance for optical lattices aiming to realize pure model systems.  The preferable approach is to define the Wannier functions in a way that the neglected processes with largest amplitude are minimized. 

\begin{figure}[t]
\includegraphics[width=\linewidth]{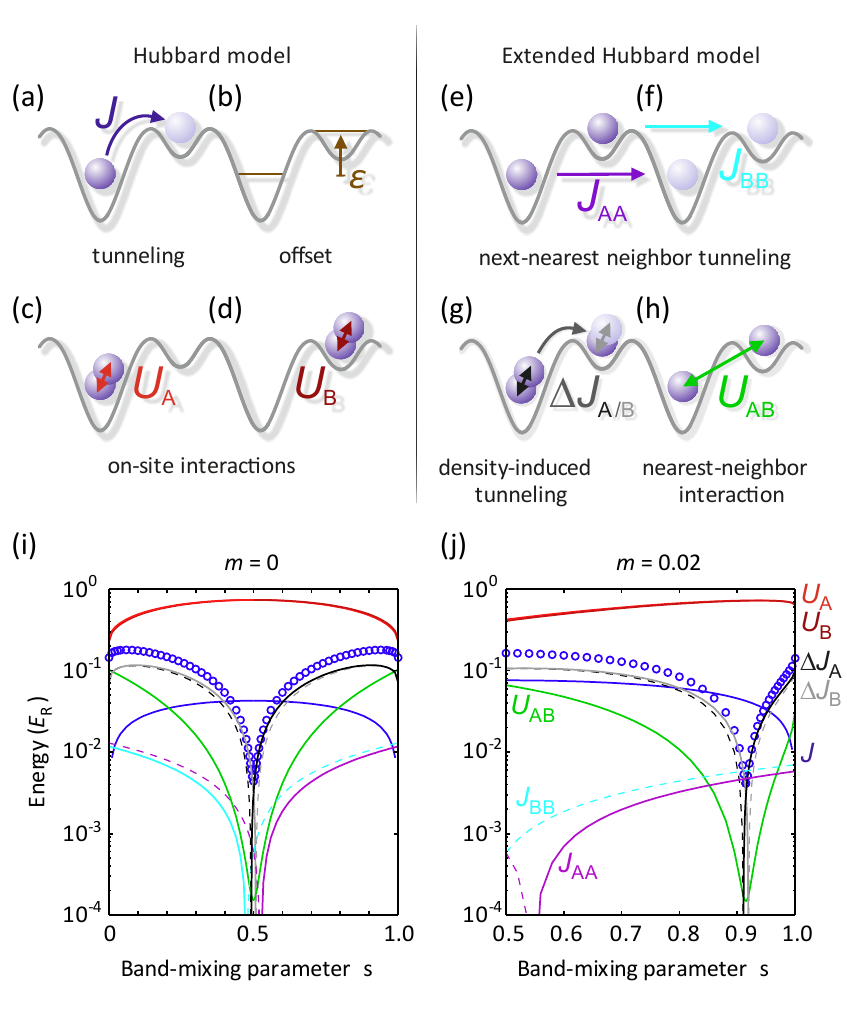}
\caption{(Color online) \textbf{(a-h)} Illustration of processes within and beyond the Hubbard model: (\textbf{a}) the tunneling matrix element $J$, (\textbf{b}) the site offset $\epsilon_B$,  (\textbf{c,d}) the next-nearest neighbor tunneling  $J_{\A\A\, /\, \B\B }$ (\textbf{e,f}) the on-site interactions $U_\A$ and $U_\B$,  (\textbf{g}) the \dit $\JBC_{\A / \B}$,  (\textbf{h}) and \dd interaction $U_{\A\B}$. \textbf{(i, j)} Parameters in dependence on the variational parameter $s$ for \textbf{(i)} $m=0$ and \textbf{(j)} $m=0.02$ at a lattice depth of $V_0=12 \ER$. Shown are the absolute values and negative signs are indicated by dashed lines. The blue circles depict the mean $(\JBC_\A^2 + \JBC_\B^2 + U^2_{\A\B})^\frac12$ which is used as the localization criterion.}
\label{SPlot}\label{Processes}
\end{figure}

The choice of the localization criterion is crucial for the validity of the resulting model. In \fig{Processes} the Hubbard processes (a-d) and the most important neglected beyond-Hubbard processes (e-h) are illustrated. The Hubbard model incorporates the tunneling $J=-\bra{w_\A} \hat H_0 \ket{w_\B}$, the site offset energy $\epsilon_\B=\bra{w_\B} \hat H_0 \ket{w_\B} -\bra{w_\A} \hat H_0 \ket{w_\A}$, as well as the on-site interaction on either site $U_{\A} \propto \int\! d^2{x}\ |w_{\A}|^4$  and $U_{\B} \propto \int\! d^2{x}\ |w_{\B}|^4$  using the Wannier function $w_{\A / \B}$ on neighboring sites $\A$ and $\B$ (see \sec{sec:TheHubbardModel}). In the case of optical lattices the \dit $\JBC_{\A/\B} \propto \int\! d^2{x}\   w_{\A}^* |w_{\A / \B}|^2 w_{\B} $ is the dominant correction of the Hubbard model \cite{Luhmann2012,Jurgensen2012,Mering2011} whereas \dd interaction $U_{\A\B}$ and next-nearest neighbor tunneling $J_{\A\A}$ and  $J_{\B\B}$ are usually small (see \sec{sec:TheExtendedHubbardModel}). Minimizing the squared sum of $\JBC_{\A/\B}$ and $U_{\A\B}$ guaranties that the system is optimally described by tunneling, on-site interactions and site offsets. Since the parameters for off-site processes are derived from the wave function overlap between neighboring lattice sites, their minimization leads to the localization of the Wannier functions on individual sites.  Note that perpendicular to the plane of the honeycomb lattice, a deep additional one-dimensional lattice ($V_z=44 \ER$) is applied, which is described by conventional one-dimensional Wannier functions for the sinusoidal potential.

For the honeycomb lattice with a bi-atomic unit cell one can construct Wannier functions from the lowest two s-bands $\sg$ and $\su$. The Wannier functions $w_\A$ and $w_\B$ on sublattices A and B can be constructed by the summation
 \begin{equation}
	\ket{w^{i}_{\A/\B}}\! =\! {\frac{1}{\sqrt{\Ns}} } \sum _{{\mathbf{k}}\in\text{BZ}}\!\! \e{ - \mathrm{i}{\mathbf{k}} \mathbf{G}_i }  \big[
			\nu^\sg_{{\mathbf{k}}, {\A/\B}}   \ket{\phi^\sg_{\mathbf{k}}}
			+ \nu^\su_{{\mathbf{k}}, {\A/\B}}  \ket{\phi^\su_{\mathbf{k}} }  \big].
 \end{equation}
over the Bloch functions of $\Ns$ reciprocal lattice vectors in the Brillouin zone (BZ). The lattice vector $\mathbf{G}_i=i_1 \mathbf{a} + i_2 \mathbf{b}$ determines the  two-site unit cell where the Wannier function is localized.

The complex coefficients $\nu$ of the Bloch functions $\ket{\phi^\sg_{\mathbf{k}}}$  and $\ket{\phi^\su_{\mathbf{k}}}$  can be written as
\begin{equation}\begin{split}
\nu^\sg_{{\mathbf{k}},\A}  =\sqrt{s}\ e^{-\mathrm{i}\theta^\sg_{{\mathbf{k}},\A}}, &\quad   \nu^\su_{{\mathbf{k}},\A}=\sqrt{1\!-\!s}\ e^{-\mathrm{i}\theta^\su_{{\mathbf{k}},\A}} \\
\nu^\sg_{{\mathbf{k}},\B}  =\sqrt{1\!-\!s}\ e^{-\mathrm{i}\theta^\sg_{{\mathbf{k}},\B}}, &\quad \nu^\su_{{\mathbf{k}},\B}=\sqrt{s}\ e^{-\mathrm{i}\theta^\su_{{\mathbf{k}},\B}} .
\end{split}\end{equation}
for sublattice A and B, respectively, with phases $\theta_{{\mathbf{k}},\A/\B}$ and a band-mixing parameter $s$. The phases must be chosen in a way that the Bloch functions at the center $\mathbf{r}_{\A/\B}$ of the respective lattice site are positive real. This constructive summation is achieved by
\begin{equation}
\theta^\n_{{\mathbf{k}},\A / \B}=\arg  \phi^\n_{\mathbf{k}} (\mathbf{r}_\text{A/B}),
\end{equation}
where $n$ denotes the band index ($\sg$ or $\su$). This is the usual procedure for the definition of Wannier functions of a one-dimensional lattice with equivalent lattice sites. The choice of $\mathbf{r}_\text{A/B}$ must obey both the orthonormality
\begin{equation}
\sum_{\mathbf{k},n} |\nu^\n_{{\mathbf{k}},\A/\B}|^2=1 \quad\ \text{and}  \quad\ \sum_{\mathbf{k},n} \nu^{\n*}_{{\mathbf{k}},\A} \nu^\n_{{\mathbf{k}},\B}=0.
\end{equation}
and spatial three-fold rotational symmetry of the Wannier functions. Due to the translational invariance with respect to the lattice vectors, it is sufficient to determine the Wannier functions  $\ket{w_\A}$ and $\ket{w_\B}$ within the unit cell at the origin $i=(0,0)$.

 \begin{figure}[t]
\includegraphics[width=\linewidth]{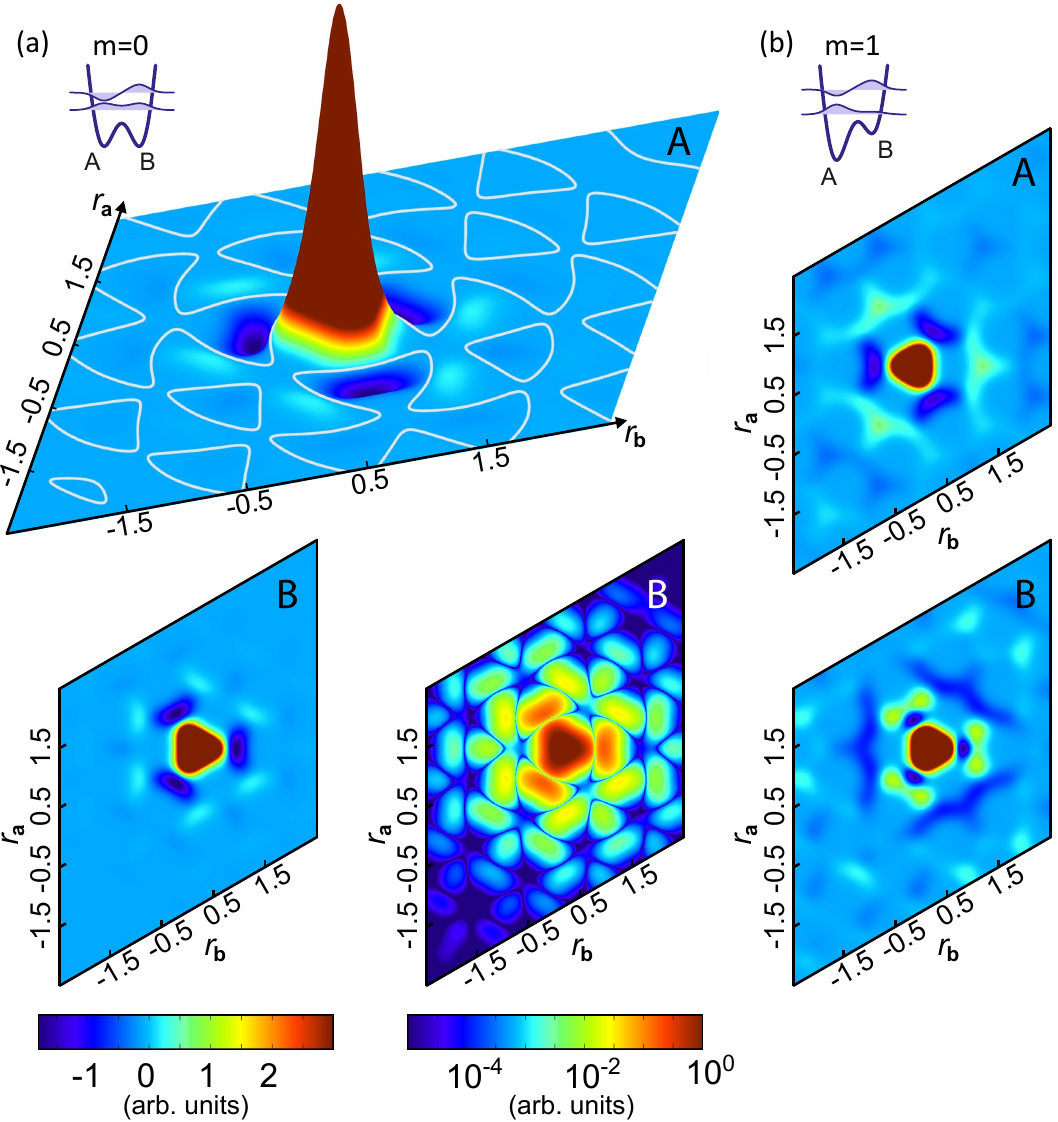}
\caption{(Color online) (\textbf{a}) The Wannier functions $\ket{w_\A}$ and $\ket{w_\B}$ for sublattices A and B at $V_0=3\ER$ for the symmetric case $m=0$. The plotted area contains $5\times5$ unit cells. The white contour lines in the upper plot for $\ket{w_\A}$ indicate the zero crossings of the Wannier function. The lower plots show $\ket{w_\B}$ on a linear and its absolute value on a logarithmic scale. Since both sublattices are identical for $m=0$, $\ket{w_\A}$ and $\ket{w_\B}$ are identical under reflection.
 (\textbf{b})  The Wannier functions for the strongly non-symmetric case $m=1$.}
\label{Wanniers}
\end{figure}

Only the single variational parameter $s$ must be optimized in order to fulfill the localization criterion. This band mixing parameter describes the individual contributions of the lowest two bands to the Wannier functions $\ket{w_\A}$ and $\ket{w_\B}$ on sublattices A and B and is crucial to obtain maximally localized Wannier functions. Figure \ref{SPlot}i and \ref{SPlot}j show the dependence of the amplitudes for all relevant processes on the variational parameter $s$ for the cases $m=0$ and $m=0.02$. For this small value of $m$, the offset energy $\epsilon$ is already on the order of the on-site energy $U$. The band mixing parameter ranges from an equal superposition with $s=\nicefrac{1}{2}$ to the limits $s\to 0$ and $s\to 1$. In the latter cases, the Wannier functions for A and B sites are constructed entirely from $\sg$ and $\su$ Bloch waves, respectively, and therefore their tunneling matrix element vanishes. Note also that the on-site interaction energy $U$ and site offset energy $\epsilon$ are relatively robust, while the tunneling $J$ and in particular off-site interaction processes are strongly influenced by the applied localization criterion. It is in general not possible to minimize the parameters of all beyond-Hubbard processes simultaneously. However, in our case the \dit processes (gray and black lines) represent the dominating corrections and their minimization leads to \textit{optimal} Wannier functions for the definition of the Hubbard model \eqref{eq:SimpleHubbardModel} restricted to $U$, $J$, and $\epsilon$. Only in the superfluid regime of shallow lattices, the next-nearest neighbor tunneling represents a significant contribution and one could use the squared sum of all neglected processes as a localization criterion. 

The resulting Wannier functions for $m=0$ and the strongly asymmetric case $m=1$ are plotted in \fig{Wanniers}. For the important case of equivalent sublattices ($m=0$) the variational parameter is simply $s=\nicefrac{1}{2}$ as follows from symmetry arguments. The logarithmic plot in \fig{Wanniers}a shows that Wannier functions are well localized even for very shallow lattices ($V_0=3\ER$). Note that very similar results are obtained using the (numerical expensive) minimization of the spread function as performed very recently for the honeycomb optical lattice in Refs.~\cite{Walters2013} and \cite{IbanezAzpiroz2013}. The situation is however more subtle for the strongly asymmetric case $m=1$. Here, the two lowest bands are not energetically well separated from higher bands (see \fig{fig:bandstructure}a). In fact, the second band is already strongly hybridized with the $\text{p}_\text{uu}$ band and hence an admixture of the forth band with a second band mixing parameter is necessary. The contribution of the $\text{p}_\text{gg}$ band is still negligible due to symmetry arguments.

In conclusion, we have presented an efficient construction scheme for Wannier functions of the honeycomb lattice. It is based on the general goal to define maximally localized Wannier functions in a way that the amplitudes of processes that are neglected in the Hubbard model are minimized. This approach results in the \textit{optimal} Hubbard model for the description of the many-body problem. Furthermore, the method is applicable to other lattice geometries with multi-atomic unit cells and is numerically inexpensive.

\subsection{The Hubbard model}
\label{sec:TheHubbardModel}

 \begin{figure*}[t]
\includegraphics[width=\linewidth]{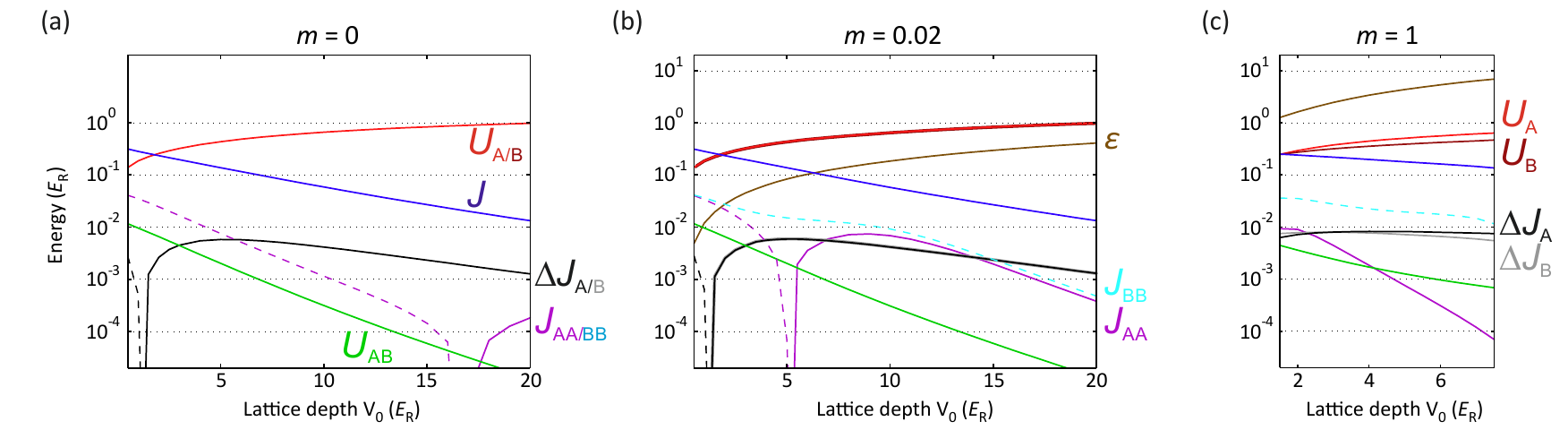}
\caption{(Color online) Hubbard and extended-Hubbard parameters for effective magnetic quantum numbers (\textbf{a}) $m=0$,(\textbf{b}) $m=0.02$, and (\textbf{c}) $m=1$. Within the Hubbard model \eqref{eq:HubbardModel}, only the on-site interactions $U_\A$ (red) and $U_\B$ (dark red), the tunneling matrix element $J$ (blue), and the site offset $\epsilon_B$ (brown) contribute. Additional processes are the \dit $\JBC_{\A / \B}$ (black/gray), the next-nearest neighbor tunneling  $J_{\A\A\, /\, \B\B }$,  and the \dd interaction $U_{\A\B}$. Shown are the absolute values and negative signs are indicated by dashed lines.}
\label{Parameter}
\end{figure*}

The standard Hubbard model using the tight binding approximation is restricted to the nearest neighbor tunneling $J$, on-site interaction $U$ and site offset energy $\epsilon$. By means of the Wannier functions the tunneling matrix elements between neighboring sites can be calculated using
\begin{equation}
	J=-\bra{w_\A} \hat H_0 \ket{w_\B}= -\frac{1}{\Ns}  \sum_{\mathbf{k},n} \nu^{\n*}_{{\mathbf{k}},\A} \nu^\n_{{\mathbf{k}},\B} E^\n_{\mathbf{k}}
\end{equation}
where $\hat H_0=\mathbf{p}^2/2m+V(\mathbf{x})$ denotes the single-particle Hamiltonian.
The on-site interaction reads
\begin{equation}
	U_{\A / \B}=g \dx |w_{\A / \B}(x_1,x_2)|^4 \, |w_\perp(x_3)|^4
\end{equation}
using the interaction parameter $g=\frac{4\pi\hbar}{m}\as$ and the Wannier function $\ket{w_\perp}$ of the perpendicular one-dimensional lattice. For concreteness, we use $^{87}\mathrm{Rb}$ parameters with a scattering length $\as\approx101\, a_0$. The energy offsets for the sublattices are given by $\epsilon_\A=0$ and
\begin{equation}
	\epsilon_{\B}= \frac{1}{\Ns}  \sum_{\mathbf{k},n}  \left( |\nu^{\n}_{{\mathbf{k}},{\B}}|^2 - |\nu^{\n}_{{\mathbf{k}},{\A}}|^2 \right)   E^\n_{\mathbf{k}}.
\end{equation}

These parameters allow to write the Hubbard Hamiltonian in the common way. Instead of labeling the sites by a unit cell vector $(i_1,i_2)$ and a sublattice index $\A$ or $\B$, it is more convenient to use a joint vector $j$. This way we can map the honeycomb lattice to a square lattice with a reduced number of bonds  (see \fig{LatticeMapping}e and f). We will use square brackets to recover the sublattice index $\A$ or $\B$ from the joint index $j$, i.e.,
\begin{equation}
	[j] \mapsto  \{\A,\B \}.
\end{equation}
Using this definition, the Hubbard Hamiltonian can be written as
\begin{equation}
	\hat H_\BHM= -J \sum_{\expect{j,j'}} \bd{j} \hat{b}_{j'}^{\phantom{\dagger}} + \frac{1}{2} \sum_j U_{[j]} \hat n_j (\hat n_j-1) + \sum_j \epsilon_{[j]} \hat n_j
	\label{eq:HubbardModel}
\end{equation}
with ${\expect{j,j'}}$ indicating the sum over nearest neighbors. The Hubbard parameter $J$, $U_\text{A/B}$, and $\epsilon=\epsilon_\B$ for the optimized Wannier function are shown in \fig{Parameter}  for the different effective magnetic quantum numbers $m=0$, $0.02$ and $1$.

In the following we discuss processes beyond the tight binding approach and the dependence of all parameters on the effective magnetic quantum number $m$.

\subsection{The extended Hubbard model}
\label{sec:TheExtendedHubbardModel}

As discussed above, several off-site processes are already neglected in the Hubbard model. These processes are illustrated in \fig{Processes}e-h. The natural question arises how important these processes are for the phase diagram of the honeycomb lattice. From \fig{Parameter} it is clear that the first order corrections to the standard Hubbard model are the \dit and next-nearest neighbor tunneling, whereas \dd interaction and pair-tunneling are negligible. The extended Hubbard Hamiltonian with first order corrections can be written as
\begin{equation}\begin{split}
	\hat H_\EBHM=& -J \sum_{\expect{j,j'}} \bd{j} \hat{b}_{j'}^{\phantom{\dagger}}  -\sum_{\expect{\expect{j,j'}}} J_{[j][j']} \bd{j} \hat{b}_{j'}^{\phantom{\dagger}} \\
		&+ \frac{1}{2} \sum_j U_{[j]} \hat n_j (\hat n_j-1) + \sum_j \epsilon_{[j]} \hat n_j \\
		& -  \sum_{\expect{j,j'}} \bd{j} ( \JBC_{[j]} \hat n_j + \JBC_{[j']} \hat n_{j'})  \hat{b}_{j'}^{\phantom{\dagger}}.\\
		\label{eq:ExtendedHubbardModel}
\end{split}\end{equation}
Here, $\expect{\expect{j,j'}}$ sums over all pairs of next-nearest neighbors with next-nearest neighbor tunneling matrix elements $J_{\A\A}$ and $J_{\B\B}$.
These matrix elements can be calculated using
 \begin{equation}
	J_{\A\A }=-\bra{w^j_{\A}}  \hat H_0 \ket{w^{j'}_{\A}}=- \frac{1}{\Ns}  \sum_{\mathbf{k},n}|\nu^{\n}_{{\mathbf{k}},{\A}}|^2 \e{i {\mathbf{k}} \mathbf{G}} E^\n_{\mathbf{k}}
\end{equation}
and analogously for $J_{\B\B}$, where $\mathbf{G}=\mathbf{G}_j-\mathbf{G}_{j'}$ is a lattice vector between neighboring unit cells. The \dit $\JBC$ stems from the two-particle interaction \cite{Luhmann2012}. It describes the tunneling of a single particle to a neighboring site induced by the interaction on either site. The process is therefore intrinsically occupation-number-dependent and scales linearly with $n_j\!+\!n_{j'}\!-\!1$.  The matrix element for \dit is
\begin{equation}
	\JBC_{\A / \B}= - g \dx  w_{\A}^*({\mathbf{r}}) |w_{\A / \B}({\mathbf{r}})|^2 w_{\B}({\mathbf{r}})    \ |w_\perp(x_3)|^4,
\end{equation}
with ${\mathbf{r}}=(x_1,x_2)$. Other processes stemming from the two-particle interaction, namely, the \dd interaction $U_{\A\B}$ and pair tunneling, are given by
\begin{equation}
	U_{\A\B}=  g \dx  |w_{\A}({\mathbf{r}})|^2 |w_{\B}({\mathbf{r}})|^2    \ |w_\perp(x_3)|^4
\end{equation}
and $J_\text{pair}=U_{\A\B}/2$. They are typically smaller than the \dit and thus are neglected here (see \fig{Parameter}). This is characteristic for the optical lattice and can be different for other potentials, e.g., solid-state materials. For the case $m=0$ with equivalent sublattices $\A$ and $\B$, the extended Hubbard parameters (\fig{Parameter}a) have qualitatively a similar dependency as for cubic lattices \cite{Luhmann2012}. It is worth noticing that the \dit $\JBC$ (black line) is about one tenth of the conventional tunneling $J$ (blue) for a wide range of parameters and can thus strongly influence the behavior of the system. For an average filling of three ($\rho=(n_\A+n_\B)/2=3$), this corresponds to an increase of the effective tunneling by $50\%$.

Furthermore, the ratio $\JBC/J$ depends on the transversal lattice depth and the scattering length. The next-nearest-neighbor tunneling couples anti-ferromagnetically and is an important contribution for shallow lattices. For intermediate and deep lattices, the next-nearest-neighbor tunneling (purple) and the \dd interaction (green) have small contributions.

As shown in \fig{Parameter}c, for $m=1$ the site offset $\epsilon_\B$ is the dominating energy. Already for relatively shallow lattices, it is larger than all other matrix elements, which causes a depopulation of the sublattice $\B$ as discussed in detail below. Another feature arising from the inequivalent sublattices is a splitting of the on-site interactions $U_\A$ and $U_\B$, \dit $\JBC_{\A / \B}$, and in particular the next-nearest neighbor tunneling $J_{\A\A}$ and $J_{\B\B}$. The absolute values of $J_{\B\B}$ are considerably larger than for  $J_{\A\A}$ due to the larger spatial extend of the $\ket{w_\B}$. This also causes the on-site interaction on $\B$ sites to be smaller than on $\A$ sites. For $m=1$, the next-nearest neighbor tunneling $J_{\A\A}$ couples ferromagnetically and  $J_{\B\B}$ anti-ferromagnetically.

\subsection{Tunable lattice site offsets}
\label{sec:TunableOffsets}

As mentioned above, $m$ can be continuously tuned by tilting the magnetic field axis with respect to the lattice plane (Eq.~\eqref{eq:m}). While the case of $m=0$ corresponds to equivalent sublattices $\A$ and $\B$, a value of $m=1$ already causes a depopulation of the sublattice $\B$. The tunability of $m$ also allows to address  the more interesting situation where the offset energy is comparable with the tunneling or the on-site interaction. The resulting competition leads to a rich phase diagram with a multitude of insulating phases with fractional fillings.

\begin{figure}[t]
\includegraphics[width=\linewidth]{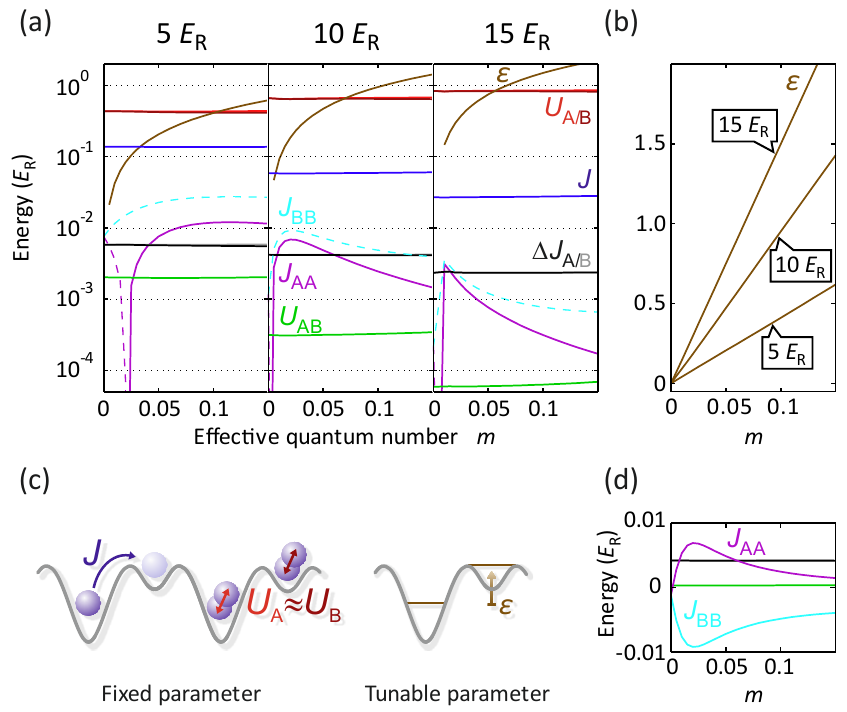}
\caption{(Color online) (\textbf{a})  The amplitude on- and off-site processes (see Figs.~\ref{Parameter} and \ref{Processes}) as a function of the effective magnetic quantum number $m$ for $V_0=5\ER$, $10\ER$, and $15\ER$.  (\textbf{b}) Non-logarithmic plot of the site offset $\epsilon$ showing its linear dependency. (\textbf{c}) Hubbard model with tunneling $J$, on-site interaction $U_\A=U_\B$, and tunable site offset $\epsilon$. (\textbf{d}) The change between ferromagnetic and anti-ferromagnetic coupling of the next-nearest-neighbor tunneling on a linear scale.}
\label{Offset}
\end{figure}

Figure \ref{Offset}b shows that the offset $\epsilon_\B$ is approximately a linear function of the parameter $m$. Importantly, the tunneling and the on-site interaction (\fig{Offset}a) have only a weak dependence on $m$ in the plotted region, i.e., $J(m)\approx J$ and $U_{\A}(m) \approx U_{\B}(m) \approx U$. Neglecting other contributions, the tuning of $m$ allows therefore to implement a Hubbard model with adjustable offset of sublattice $\B$ as illustrated in \fig{Offset}c. The model Hamiltonian can be written as
\begin{equation}\begin{split}
	\hat H_\BHM(m)=& -J \sum_{\expect{j,j'}} \bd{j} \hat{b}_{j'}^{\phantom{\dagger}} + \frac{1}{2} \sum_j U \hat n_j (\hat n_j-1)\\
		&+ \sum_j \epsilon_{[j]}(m) \hat n_j.
		\label{eq:SimpleHubbardModel}
\end{split}\end{equation}
As an important result, in the spirit of quantum simulation, the tuning of the effective quantum number $m$ represents an additional tool for engineering many-body Hamiltonians.

In the case of equivalent sublattices ($m=0$), the \dit is the leading order correction of the model Hamiltonian \eqref{eq:SimpleHubbardModel}, whereas for larger values of $m$ the next-nearest neighbor tunneling can have a similar amplitude (\fig{Offset}d).  We determine the impact of these additional processes in \sec{sec:PhaseDiagrams}.

\section{Phase diagrams}
\label{sec:PhaseDiagrams}

\begin{figure*}[t]
\includegraphics[width=\linewidth]{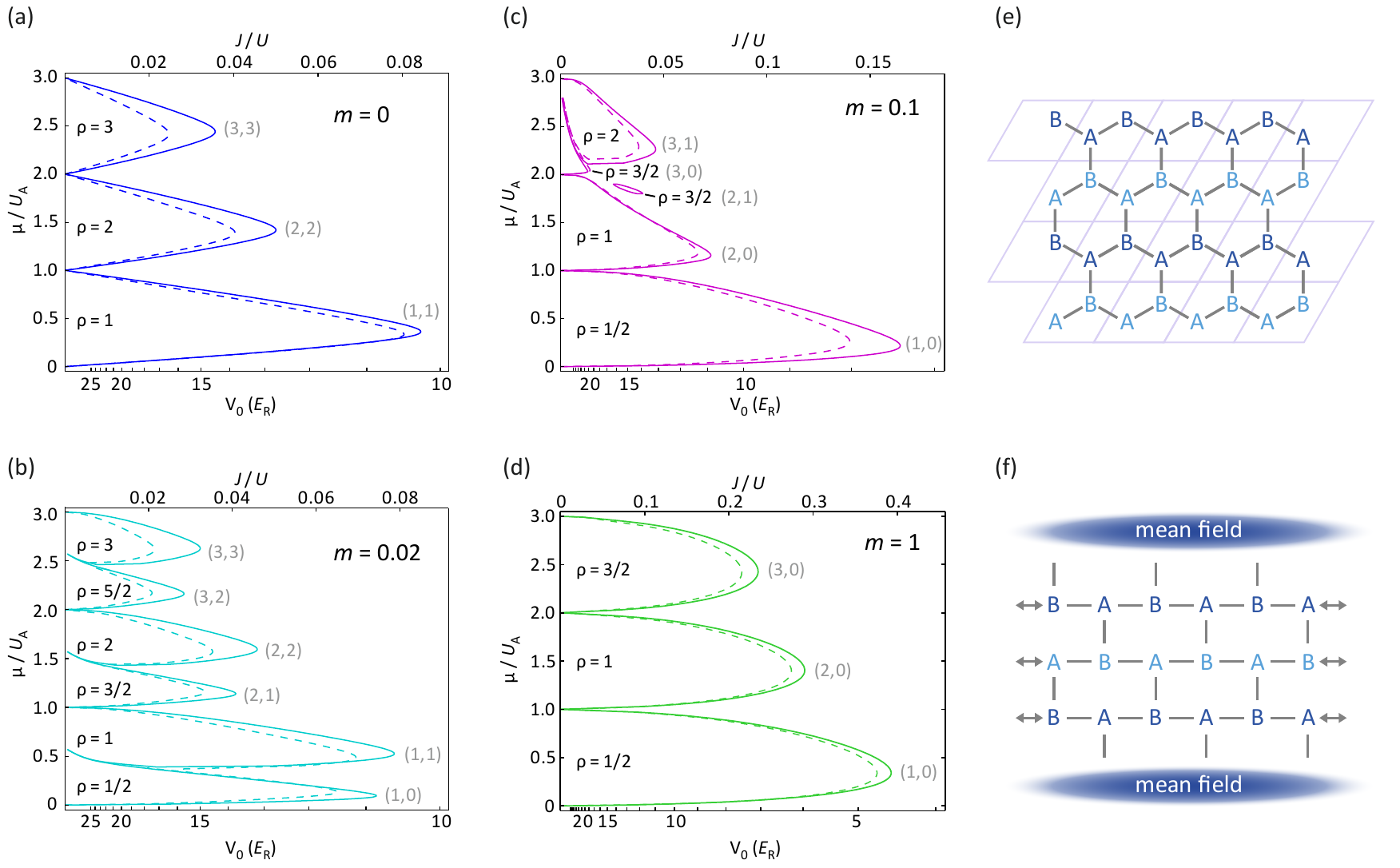}
\caption{(Color online) (\textbf{a-d}) Superfluid to Mott insulator transition  for $m=0$, $0.02$, $0.1$, and $1$, where $\mu$ is the chemical potential. The Mott lobes are shown as a function of the lattice depth $V_0$ (axis at the bottom) and the ratio $J/U$ (top). Mott phases with integer and half-integer filling $\rho$ exist for non-equivalent sublattices $\A$ and $\B$ with the occupation numbers $(n_\A,n_\B)$ (gray). The results for the Hubbard model \eqref{eq:HubbardModel} are shown as  solid lines and for the extended model \eqref{eq:ExtendedHubbardModel}  as dashed lines. The phase boundary within the Hubbard model depends only on the ratio $J/U$ assuming $U=U_{\A}\approx U_{\B}$. (\textbf{e}) Lattice structure of the honeycomb lattice and (\textbf{f}) mapping  on a square lattice with reduced bonds. The depicted cluster of sites is used for the cluster Gutzwiller calculation, where the double arrow indicates periodic boundary conditions.}
\label{LatticeMapping}\label{SFMIHom}
\end{figure*}

In the following, the phase diagrams of bosonic atoms in the tunable state-dependent honeycomb lattice are discussed. We consider the general case of  Hubbard models for hexagonal lattices  superimposed with a bi-atomic superlattice structure with arbitrary site offsets. Mean-field calculations in Ref.~\cite{SoltanPanahi2011} only allow for approximative results due to the small number of nearest-neighbors in the honeycomb lattice \cite{Luhmann2013,Teichmann2010}. Therefore, we apply a bosonic cluster mean-field approach \cite{Buonsante2004b, Buonsante2005b, McIntosh2012, Pisarski2011, Jain2004, Yamamoto2012, Yamamoto2012b, Hen2009, Hen2010, Luhmann2013}. It has been shown that this approach gives accurate results for the Bose-Hubbard model for hexagonal lattices \cite{Luhmann2013}. Furthermore, we discuss the phase diagrams for experimental parameters in dependence on the effective magnetic quantum number $m$ and the influence of extended Hubbard processes such as \dit and next-nearest neighbor tunneling. Finally, by using the improved Wannier functions from Section \ref{sec:DefWannier} and accurate numerics to calculate the phase diagram we show that the theoretical predictions match very well with the experimental results in Ref.~\cite{SoltanPanahi2011}.

\subsection{Cluster Gutzwiller method}

We briefly review the cluster Gutzwiller method applied in the following. The idea is to solve the many-particle problem for a cluster of lattice sites which is coupled to the mean-field at its boundary \cite{Buonsante2004b, Buonsante2005b, McIntosh2012, Pisarski2011, Jain2004, Yamamoto2012, Yamamoto2012b, Hen2009, Hen2010, Luhmann2013}. The exactly treated cluster is decoupled from the surrounding lattice by replacing all operators that act on sites outside the cluster with their expectation values. In a self-consistent procedure, the mean-field is determined from the solution of the previous iteration. This is a natural extension of the conventional Gutzwiller approach, where a single lattice site is decoupled from the lattice. The striking advantage is that intersite correlations can be captured, enhancing the precision significantly and giving access to correlated quantum phases such as so-called loophole insulators \cite{Buonsante2004b,Buonsante2005}. For the extended Hubbard model, the cluster Gutzwiller method requires the two different mean-field parameters $\expect{\hat b}$ and $\expect{\hat n\hat b}$, where the latter is introduced by the density-induced tunneling process.
 
In the many-particle cluster basis $\ket{N}$ the Hamiltonian matrix elements
\begin{equation}
	\hat H_{MN}=\bra{M} \hat H_\text{cluster}   + \hat H_\text{boundary}  \ket{N}
	\label{eq:HamiltionanMatrix}
\end{equation}
decompose in two parts describing the cluster and its boundary. For the general case of the extended Hamiltonian \eqref{eq:ExtendedHubbardModel}, we have $\hat H_\text{cluster}= \hat H_\EBHM - \mu \sum_j \hat n_j$, where $\mu$ is the chemical potential. The Hamiltonian $\hat H_\text{boundary} $ describes the coupling of sites at the boundary of the cluster to sites outside the cluster. For an infinite system, we can obtain the expectation values for sites outside the cluster from two innermost sites in the cluster, i.e., \textit{target} sites $\text{a}$ and $\text{b}$ of the sublattice $\A$ and $\B$, respectively. Consequently, a site $j$ of sublattice $\A$ (B analogously) couples via
\begin{equation}\begin{split}
	\hat H_\text{boundary} ^j =&-J\  \nu_j\ \bd{j} \expect{\hat{b}_{b}^{\phantom{\dagger}}}   - J_{\A\A}\ \nu^{\text{NN}}_j\ \bd{j} \expect{\hat{b}_{b}^{\phantom{\dagger}}} \\
  &- \JBC_\A\ \nu_j\ \bd{j} \hat n_j   \expect{\hat{b}_{b}^{\phantom{\dagger}}}   - \JBC_\B\ \nu_j\ \bd{j} \expect{\hat n_b\hat{b}_{b}^{\phantom{\dagger}}} \\
  & +  c.c.    ,
\end{split}\end{equation}
where  $\nu_j$ ($\nu^{\text{NN}}_j$) denotes the number of nearest (next-nearest) neighbors outside the cluster. For the Bose-Hubbard model we have $J_{\A\A}=\JBC_\A=\JBC_\B=0$ and the expression above simplifies drastically. For the calculations, we use a cluster of $18$ sites as shown in \fig{SFMIHom}f, where periodic boundary conditions are applied along the horizontal direction (see Ref.~\cite{Luhmann2013} for further details). The latter reduces the number of bonds to the mean-field, where in this case the finite-size scaling parameter measuring the ratio of internal cluster bonds to total bonds is  $\lambda=0.8$ ($\lambda=0$ corresponds to a single-site, $\lambda=1$ to an infinite cluster).

\begin{figure*}
\includegraphics[width=\linewidth]{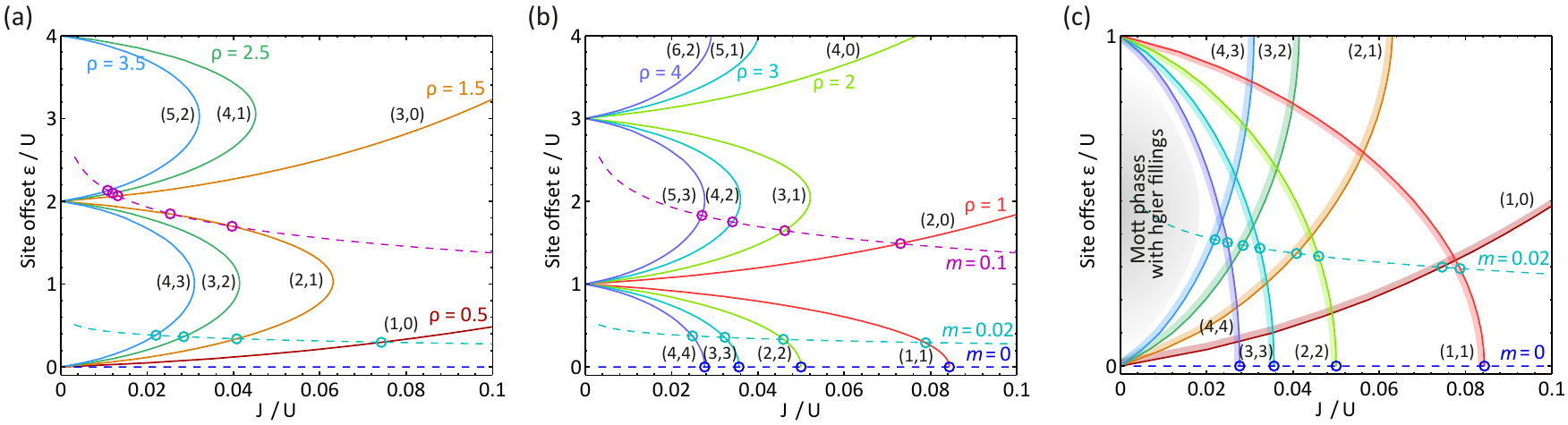}
\caption{(Color online) Universal  phase diagram for arbitrary values of  $m$ in the ${\epsilon}/{U}$--${J}/{U}$ plane for (\textbf{a}) half-integer and  (\textbf{b}) integer filling factors up to $\rho=4$. The intersections of the Mott lobes with the function ${\epsilon}/{U}$ for different values of $m$ (dashed lines) corresponds to the SF-MI transition point at the tip of the Mott lobe in \fig{SFMIHom}.  (\textbf{c}) Phase diagram combining half-integer and integer filling factors for site offsets $0<\epsilon<U$.}
\label{EpsilonLobes}
\end{figure*}

\subsection{State-dependent phase diagrams}
The quantum phase diagram of the state-dependent honeycomb lattice strongly depends on the effective quantum number $m$. The general feature is the transition from a superfluid state in shallow lattices (large values of $J/U$) to strongly correlated Mott-insulating states in deep lattices. The latter is defined by vanishing superfluid order parameters $\expect{\hat{b}_{a}^{\phantom{\dagger}}}=\expect{\hat{b}_{b}^{\phantom{\dagger}}}=0$ and fixed particle numbers per unit cell $(n_\A,n_\B)$. For the state-independent case $m=0$, the insulating phases are characterized by the same integer occupation of both sublattices A and B, i.e., $n_\A=n_\B=\rho$. The site-offset energy $\epsilon$ introduces an imbalance between the two sublattices, leading to insulating phases with uneven fillings $n_\A \neq n_\B$.

In \fig{SFMIHom}a-d the phase diagrams are shown for different effective magnetic quantum numbers $m$ in dependence on the lattice depth $V_0$ and the chemical potential $\mu/U_\A$. The Mott lobes are shown for the standard Hubbard model Eq.~\eqref{eq:SimpleHubbardModel} (solid lines) as well as for the extended Hubbard model Eq.~\eqref{eq:ExtendedHubbardModel} including off-site processes (dashed lines). The \dit increases the total tunneling energy and thereby shifts the transition to deeper lattices. This is in particular strong for higher filling factors which reflects the occupation-dependency of the \di tunneling. The results of the cluster method for $m=0$ differ strongly from those of the conventional mean-field approach (see Refs.~\cite{Fisher1989, Krauth1992,Oosten2001}) predicting $J/U=0.0572$ for the lowest Mott lobe with $(n_\A,n_\B)=(1,1)$. The large discrepancy is caused by the small number of nearest neighbors in the honeycomb lattice. In the case of the standard Hubbard model, the phase diagrams depend only on the ratio $J/U$ plotted at the top of each figure. However, other parameters such as site offsets or \dit also depend on the effective magnetic quantum number $m$. 

The possibility of tuning the site offset $\epsilon=\epsilon_\B$ as described in \sec{sec:TunableOffsets} leads to an interesting competition of site offset and on-site interaction. When the site-offset exceeds the on-site energy, a population imbalance is imprinted on each of the unit cells \cite{Buonsante2004,Buonsante2004b,Rousseau2006,Buonsante2005,Chen2010}. For  $m=0.02$ (\fig{SFMIHom}b), where the site offset $\epsilon_\B$ is on the order of the $U/2$,  this criticality is reflected by alternating Mott lobes with half-integer and integer filling $\rho$, e.g., the Mott states $(1,0)$ and $(1,1)$. Depending on the chemical potential $\mu$ transitions to both Mott configurations from the superfluid are possible. In \sec{sec:Universal}, we discuss in detail how the site offset $\epsilon_ \B$ affects the Mott transition and how the critical point can be determined for a given offset. In deep lattices, the boundaries between both Mott phases are strongly  bent and separated by a very narrow superfluid region. The latter is triggered by the increase of $\epsilon_\B$ with the lattice depth (see \fig{Parameter}b). 

By increasing the effective magnetic quantum number ($m=0.1$ and $m=1$), a depopulation of the $\B$ lattice sites occurs due to the large energy offsets. 
For $m=0.1$, the phase diagram for the higher Mott lobes is rather complex and surprisingly the $(2,1)$ Mott phase is completely surrounded by the superfluid phase, which is further elaborated in \sec{sec:Universal}. In the extended Hubbard model, where the \dit causes in general smaller Mott phases, the $(2,1)$ Mott insulator is not a ground state for $m=0.1$. For the case $m=1$, corresponding to $\mF=1$ atoms at a perpendicular magnetic field, only $(n,0)$ Mott insulator phases can be observed (\fig{SFMIHom}d). However, in the superfluid phase the $\B$ sites are nonetheless important as they induce the fluctuations between the $\A$ sites via second order tunneling on the order of $J^2/\epsilon_\B$. Direct next-nearest neighbor hopping between the $\A$ sites contributes only to a minor degree which can be deduced from the small difference between standard and extended Hubbard model (solid and dashed lines). Due to the large value of $\epsilon_\B$ the Mott transition to $(1,0)$ occurs at high values of $J/U\approx 0.4$ and therefore already in very shallow lattices.

\subsection{Universal $\epsilon$--$J$--$U$ phase diagram}
\label{sec:Universal}

While in \fig{SFMIHom} the phase diagrams for specific values of effective magnetic quantum number $m$ are discussed, we show in the following the results for the standard Hubbard model \eqref{eq:SimpleHubbardModel}
\begin{equation*}
	\hat H_\BHM= -J\sum_{\expect{j,j'}} \bd{j} \hat{b}_{j'}^{\phantom{\dagger}} + \frac{U}{2} \sum_j \hat n_j (\hat n_j-1)
	+ \sum_j \epsilon_{[j]} \hat n_j
		\label{eq:SimpleHubbardModelRepeated}
\end{equation*}
in dependence on the site offset $\epsilon$. The positions of the tips of the Mott lobes in \fig{SFMIHom} are of particular interest, since they mark the transitions into the insulating phases at (half) integer filling. Fixing the chemical potential $\mu$ to the corresponding value allows to draw the universal phase diagram in the $\epsilon/U$ -- $J/U$ plane as depicted in Fig.~\ref{EpsilonLobes}. In this representation, lobes of Mott phases $(n_\A,n_\B)$ with filling $\rho=(n_\A+n_\B)/2$ emerge. Mott phases with a given imbalance $\Delta n=n_\A-n_\B$ exist for $n_\B>0$ in a range $\Delta n-1 < \epsilon/U < \Delta n+1$ reflecting the competition between the on-site energy $U$ and offset energy $\epsilon$.

The dashed lines in \fig{EpsilonLobes} represent the site offset $\epsilon/U$ for different values of $m$ as a function of $J/U$. For specific $m$, the superfluid to Mott insulator transitions with $(n_\A,n_\B)$ particles are given by the intersection with the respective Mott lobe (open circles). They correspond to the tips of the lobes for $m=0$, $0.02$, and $0.1$ in Figs.~\ref{SFMIHom}a-c, where $\epsilon$ is solely determined by $m$ and the optical lattice potential. In \fig{EpsilonLobes}c, where even and odd fillings are plotted, we can determine the critical values for $m=0$ and $m=0.02$. It is clear that for $m=0$ only transitions to $(n,n)$ Mott phases exist, whereas for $m=0.02$ both $(n,n)$ and $(n,n-1)$ lobes can be found with comparable transition points (see \fig{SFMIHom}b). In \fig{EpsilonLobes}a the ratio $\epsilon/U$ for $m=0.1$ (purple line) has two intersections with the $(2,1)$ Mott lobe, which indicates that this phase only exists in the range of $0.026 \lesssim J/U \lesssim 0.04$. For lower values of $J/U$ the site offset $\epsilon/U$ increases and the superfluid phase is reentered. As a result a Mott insulator island appears in the phase diagram in \fig{SFMIHom}c. Note that for the extended Hubbard model the insulator phases are in general smaller and the $(2,1)$ Mott phase only appears for values $m<0.1$.

In conclusion, the representation in \fig{EpsilonLobes} allows us to predict the possible Mott phases for arbitrary site offsets by drawing the line ${\epsilon}/{U}$. This ratio is determined by the Wannier function computed in \sec{sec:BandStructureAndWannierFunctions} for given values of $m$ and $V_0$. The critical values $J/U$ for entering the Mott phases (and possibly reentering the superfluid phase) are given by the intersections of the phase boundaries with the respective line $\epsilon/U$.

\begin{figure}[t]
\includegraphics[width=\linewidth]{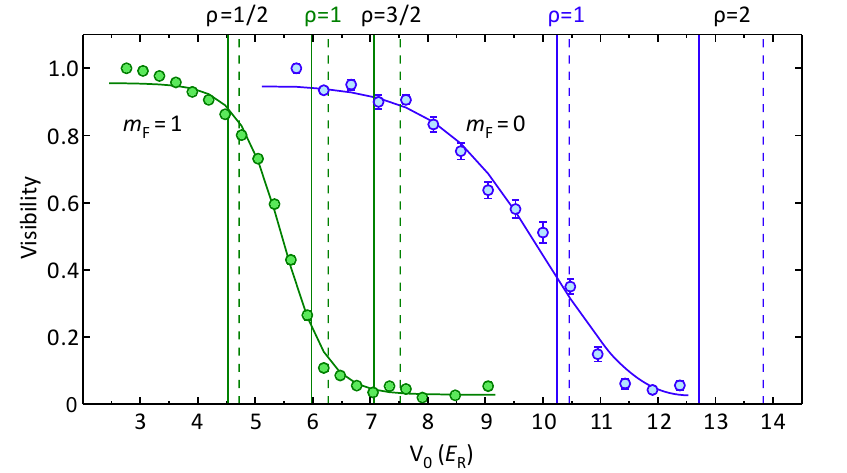}
\caption{(Color online) Comparison with the experimental data from Ref.~\cite{SoltanPanahi2011} which depicts the visibility after time of flight (circles) for $m=0$ (blue) and $m=1$ (green). The vertical lines are the theoretical predictions for the Mott insulator transition for filling factor $\rho$ in both cases. The solid lines show the result for the Hubbard and the dashed lines for the extended model.  The experimental data is in good agreement with transitions at a filling factor of $\rho=1$.}
\label{Exp}
\end{figure}

\subsection{Comparison with experimental data}

In \fig{Exp} we compare the theoretical predictions for $m=0$ and $1$ with the experimental data in Ref.~\cite{SoltanPanahi2011}. In this experiment,  the described state-dependent honeycomb lattice was realized and loaded with $^{87}$Rb atoms in the hyperfine ground-state manifold $F=1$ and $F=2$. This allows to study the superfluid to Mott insulator transition with different magnetic quantum numbers $m_F$. However, without the proposed rotation of the magnetic field axis only integer effective magnetic quantum numbers are accessible. The experimental data in \fig{Exp} shows the visibility of the atomic cloud after time-of-flight expansion, which vanishes in the Mott insulator phase. However, in the experiment the additional confinement leads to a slowly decreasing local chemical potential from the trap center, which increases the overall visibility close to the Mott transition due to the coexistence of Mott plateaus and superfluid rings.

Our calculation for the critical points is plotted as vertical lines, where the solid lines depict the standard Hubbard and the dashed lines the extended model. For $m=0$, the critical value $V_0^\text{c}\approx10.5\ER$ for the filling $\rho=1$ matches well with the experimental result (blue line). As discussed above for filling $\rho=1$ the correction in the extended Hubbard model is relatively small. For $\rho=2$, the correction with about $1\ER$ is much larger, since the \dit as the leading-order correction scales with $2\rho-1$. Note that the previous theoretical prediction \cite{SoltanPanahi2011} is $V_0\approx13\ER$-$19\ER$ for a filling $\rho=1$-$2$ is much larger and does not agree well with the measurement.

In contrast, the same calculation \cite{SoltanPanahi2011} for $m=1$ predict a much lower value than in experiment, namely $V_0\approx4\ER$-$5\ER$ for $\rho=1$-$2$ although the single-site Gutzwiller approach should overestimate the critical lattice depth \cite{Luhmann2013}. Our approach for Wannier functions in combination with the cluster Gutzwiller method predicts the transition at about $6\ER$ for $\rho=1$, which agrees well with the experimental data. Thus, the results for $m_F=0$ and $m_F=1$ both indicate an average filling of $\rho=1$.
 
\section{Conclusions}
We have presented a versatile setup for the generation of optical lattices with three-fold symmetry where the polarization of the light and an external magnetic field can be used to realize a manifold of lattice geometries. This offers promising opportunities for new optical lattice setups and grants access to completely new quantum physics. For the case of the honeycomb lattice, a tunable site-offset energy between the sublattices A and B introduces a new degree of freedom to engineer more complex lattices topologies. Here, the precise knowledge of system parameters -- Wannier functions and contributing interaction processes -- is essential for the interpretation of the experimental results. Using the cluster mean-field method and a well-suited localization criterion for the Wannier states, we have been able to compute accurate phase diagrams. The cluster mean-field method has proven to be an efficient and precise tool for the determination of the phase diagrams of both the standard and the extended Hubbard models, especially for lattices with small coordination numbers. With our results we were able to pinpoint the influence of beyond-Hubbard processes, i.e., \dit and next-nearest neighbor tunneling. A universal representation of the phase diagrams for arbitrary site offset energies has been introduced. In general, the presented results provide a major improvement on previous theoretical predictions and show excellent agreement with the experimental data in Ref.~\cite{SoltanPanahi2011}. We find that the next-nearest neighbor tunneling does not appear to cause the large discrepancy between experiment and theory in \cite{SoltanPanahi2011} as proposed. The presented methods, especially the efficient construction scheme for \textit{optimal} Wannier states, can be easily extended to other lattice geometries.

We acknowledge funding by the Deutsche Forschungsgemeinschaft (grants SFB~925 and GRK~1355).

\end{document}